\documentclass{iaus}
\usepackage{graphicx}

\title[Highly Ionised Gas as a Diagnostic of the Inner NLR] %% give here short title %%
{Highly Ionised Gas as a Diagnostic of the Inner NLR}

\author[M. Ward et al.]   %% give here short author list %%
{M.J.~Ward$^1$,
J.~Mullaney$^1$,
C.~Jin$^{1}$,
R.~Davies$^2$
}

\affiliation{$^1$Department of Physics,Durham University,
  Durham, England\\
$^2$Max-Planck Institut f\"ur extraterrestrische Physik,Garching, Germany}

\pubyear{2009}
\volume{267}  %% insert here IAU Symposium No.
\pagerange{119--126}
\jname{Co-Evolution of Central Black Holes and Galaxies}
\editors{B.M.\ Peterson, R.S.\ Somerville, \& T.\ Storchi-Bergmann, eds.}
\begin{document}

\maketitle

\begin{abstract}
The spectra of AGN from the ultraviolet to the near infrared, exhibit
emission lines covering a wide range of ionisation states, from neutral
species such as [O\,I] 6300 \AA, up to [Fe\,XIV] 5303 \AA. Here we report
on some recent studies of 
the properties of  highly ionised lines (HILs), plus two case studies
of individual objects. Future IFU observations at high spatial and good 
spectral resolution, will probe the excitation and kinematics of the gas in
the zone between the extended NLR and unresolved BLR. Multi-component
SED fitting can be used to link the source of photoionisation with the
strengths and ratios of the HILs.

\end{abstract}

\firstsection % if your document starts with a section,
\section{Introduction}

The detection and initial investigations of highly ionised emission 
line species in AGN, has been possible for several decades eg. \cite{Pel81} and
\cite{Pen84}. However it is only relatively recently that sensitive near-IR
spectroscopy and
IFUs have renewed the impetus to use these features as diagnostics of both  
the ionisation processes and gas kinematics. At high energies eg. X-rays and
the ultraviolet, observational evidence has been accumulating in support of
the presence of outflows. Although estimates of the mass in the outflow are
difficult to make and 
assumption dependent, in some cases it may be substantial, and the 
associated kinetic energy a significant fraction of the bolometric luminosity
(\cite[Reeves et al. (2009)]{Ree09}).   

\section{Samples of AGN with Highly Ionised Lines}

Thanks to advances in the technology of near-IR arrays, we 
now have significant numbers of AGN with high quality spectra from 1-2.5$\mu$
eg. \cite{Rif06} and \cite{lan08}. Although generally the near-IR spectral
resolution is not yet comparable with similar samples at optical wavelengths.
Based mostly on optical studies there is now convincing evidence of blue-shifts
of the highly ionised species with respect to the systemic velocity,
see \cite{Rod06} and
\cite{Mul08}. In a recent study \cite{Gel09} used spectra mined from the
SDSS catalogue to select 63 AGN, based on the strength of their
[Fe\,X] 6374 \AA
line emission. Clearly such a sample has a strong selection effect, but it is
valuable to compare its properties with other studies which almost aways draw
their targets from the optically brighest and best observed AGN. \cite{Gel09}
find strong correlations between the X-ray continuum and lines of
[Fe\,X] and [Fe\,XI], as well as a trend for the broadest HIL profiles to have
the highest blue-shifts. 

Smaller samples of AGN have been observed via 
narrow band imaging in the emission line of [Si\,VII] 2.48$\mu$.  
\cite{Pri05} showed
that in a small sample of Seyfert 2s, the region is extended from a few tens
up to 150\,pc, making it 
significantly less extended than the NLR. Furthermore, several studies
of the HIL, for example recent work by \cite{Mul08}, have shown that the 
profiles of the HILs can often be deconvolved into several components,
and that the broadest of these has a FWHM in between those of the BLR
and the NLR (as typified by the [O\,III] line at 5007 \AA). Finally, a very 
recent study of HILs using HST+STIS data(\cite[Mazzalay et al. 2010]{Maz10}),
confirm substantial blue-shifts, and also detect line splitting at the 
core of some galaxies.

\begin{figure}[t]
\begin{center}
\includegraphics[width=13cm,clip]{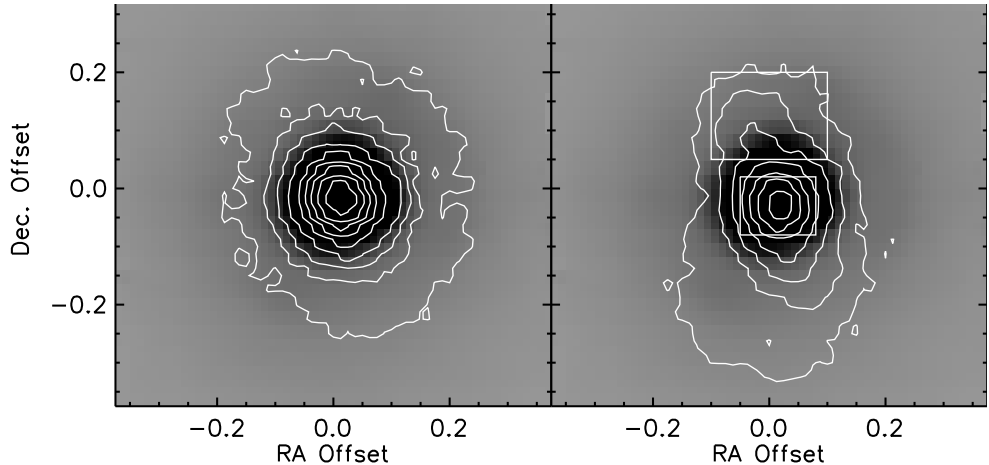}
\end{center}
\caption{NGC3783, SINFONI data: contours of broad Br$\gamma$ Flux,
assumed to be unresolved,
(left), contour of [Si\,VI] emission line at 1.963$\mu$, (right).  
Mullaney et al., in preparation (2010)}
\label{ngc3783a}
\end{figure}

\section{Case Studies}

Although statistical studies of  high ionisation line species detected in
large samples of AGN, such as those mentioned above are very useful, these 
need to be complemented by detailed studies of some individual cases.
In their seminal study of the optical and near infrared 
coronal lines in NGC\,1068, \cite{Mar96}, demonstrated the diagnostic 
potential of combining several ratios of HIL of various elements 
and ionisation species . However, NGC\,1068 is a 
Seyfert 2, and presumably some proportion of the region emitting the highly 
ionised gas will be blocked by the torus. It would therefore be desirable
to make a detailed studies of Seyfert 1s.

\subsection{Akn\,564}

 \cite{Mul09} used new spectroscopic
data on the NLS1 Akn\,564, and extended the modeling techniques described in
\cite{Fer97}. They were able to build a self consistent model which
explains both the line ratios and kinematics of the highly ionised gas. 
In essence this model invokes gas liberated from the the dusty torus, and then
accelerated to its terminal velocity. In principle this model can be tested 
and constrained using additional HIL species, and high spatial resolution 
spectroscopy.

\begin{figure}[t]
\begin{center}
\includegraphics[width=11cm]{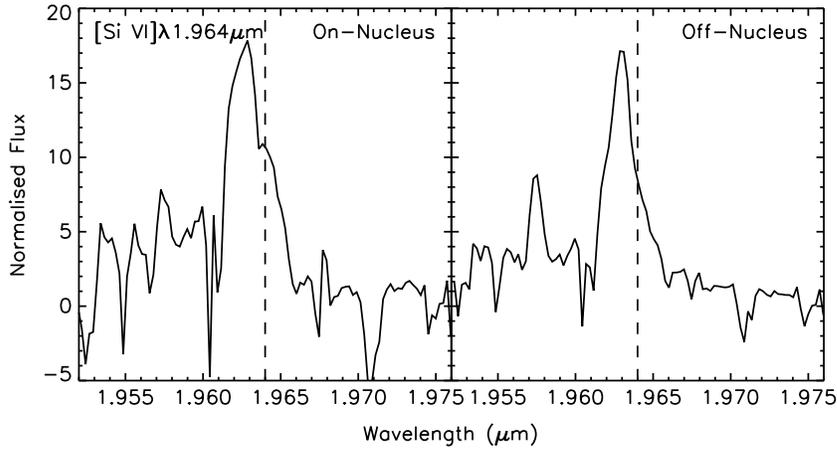}
\end{center}
\caption{NGC\,3783: [Si\,VI] emission from nucleus on (left), [Si\,VI] off-nucleus
on (right).
 Dashed line shows velocity of 1-0\,S(1) H$_2$ emission line, which is also
consistent with that of HI 21cm radio line}
\label{ngc3783b}
\end{figure}

\subsection{NGC\,3783}

 NGC\,3783 is a classic Seyfert type 1, and due to its proximity and
 high flux, it has been intensively studied at all energies. 
 We have used recent near-IR data from SINFONI, to extract maps
 of several emission lines in the K-band. Of particular interest in this 
 context is the map of [Si\,VI] at rest wavelength 
 1.963$\mu$ (Fig.~\ref{ngc3783a}). An elongation can be seen
towards the North (top), of a few tenths of an arcsec. The projected physical
extent is $\sim$42\,pc. Interestingly the blue-shift from systemic velocity
of both the on and off-nucleus profile is about the same
ie. $\sim$250\,km\,s$^{-1}$ (Fig.~\ref{ngc3783b}). This lack of a velocity
gradient is what might be expected if one were viewing a conical outflow
roughly aligned along the line of sight. More detailed modelling of 
spatially extended emission line ratios is in progress.

\begin{figure}[t]
\begin{center}
\includegraphics[scale=0.5]{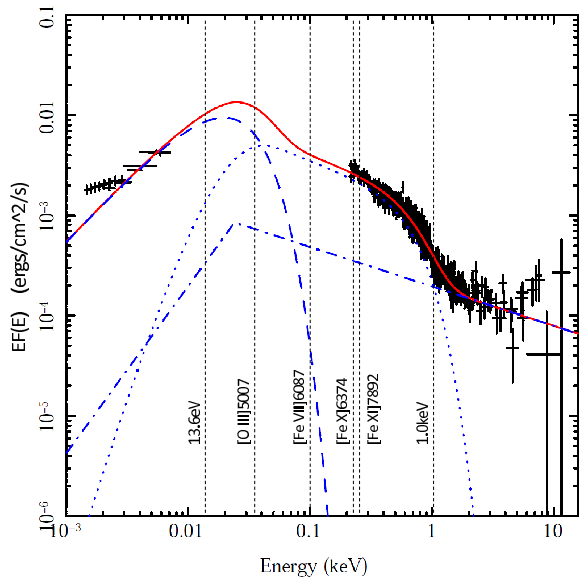}
\end{center}
\caption{Example of an AGN SED fitted using three components,disc =
dashed line, Compton component = dotted line, and broken power-law. Note:
the poor fit at lowest energies is due to stellar contamination}
\label{SED}
\end{figure}

\begin{figure}[t]
\begin{center}
\includegraphics[angle=90,width=13cm]{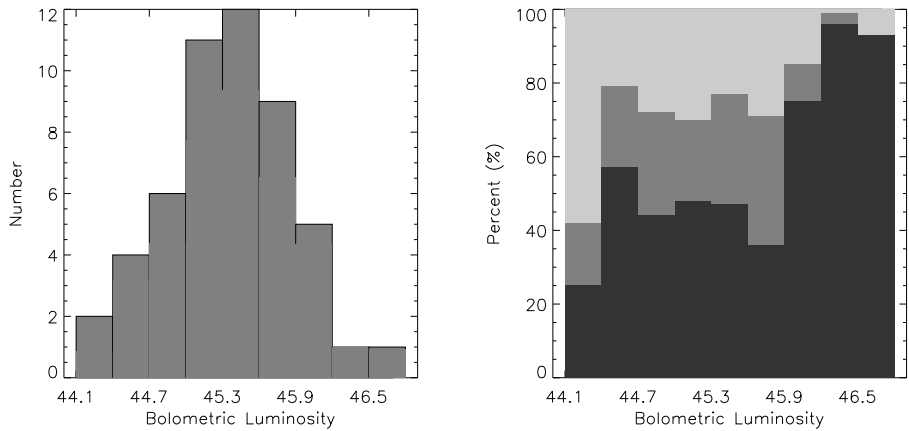}
\end{center}
\caption{Histrograms of a new X-ray sample of 51 AGN, bolometric 
luminosity (left).
On the right is the percentage contributed by each SED component,
black = disc, dark grey = Compton component, grey (top) = power-law.}
\label{hist}
\end{figure}

\section{Future Observational Tests}

Because the HIL emitting region is very probably statified, good spectral
resolution is essential to deconvolve the line profiles. This, combined with
high spatial resolution is now possible using IFUs such as SINFONI, and 
a start has been made on a few nearby AGN. Other routes are also available.
For example, one of the key open questions regarding the innermost
HIL region, is the nature of the ionisation mechanism. This is assumed to
be via photionisation
based on estimates of the gas temperature, and good correlations with the high
energy continuum. However, the crucial continuum photon energies lie in the 
unobservable region of the spectrum. To circumvent this, we are analysing a 
sample of 51 AGN all with excellent X-ray data, plus SDSS spectra. We 
have fitted their SEDs using three principal components; an accretion disc,
a Compton component, and a broken power-law, see Fig.~\ref{SED} for one
example. We shall be able to use the integrated luminosities of each 
continuum component Fig.~\ref{hist},
to see which correlates best with the HILs. Using 
black hole mass estimates from the Balmer line profiles and continua,
we shall see whether there is any relationship between the Eddington
ratio and the strength,velocity width and blue-shift of the HILs.

\end{document}